\documentstyle[12pt]{article}
\input tcilatex

\begin{document}

\author{S. L. Lyakhovich, A. A. Sharapov and K. M. Shekhter}
\title{MASSIVE SPINNING PARTICLE \\
IN ANY DIMENSION. \\
II. (HALF-)INTEGER SPINS}
\date{{\it Physics Department, Tomsk State University, 634050 Tomsk, Russia }}
\maketitle

\begin{abstract}
The general model of an arbitrary spin massive particle in any dimensional
space-time is derived on the basis of Kirillov - Kostant - Souriau approach.
\end{abstract}


\section{Introduction}

The Lagrangian description of the relativistic spinning particles is one of
the recurrently discussed themes in high energy physics, having a long
history. The retrospective exposition of the question and some basic
references can be found in the review paper \cite{Frydryszhak}. In the
context of $M$-theory the models of spinning particles awake today fresh
interest as special but highly nontrivial examples of $0$-branes which along
with some other extended objects are considered to be the basic ingredients
of the non-perturbative string theory \cite{Townsend}. Since the target
space of a consistent string theory has certainly more than four dimensions (
$10$ or perhaps $11$) the observable space-time is supposed to result from
the Kaluza-Klein compactification of extra dimensions in the low energy
limit. In so doing the effective dimension of p-brane may decrease down to
zero when it is considered from the viewpoint of four-dimensional observer.
The classical example is the double dimensional reduction of $d=11$
supermembrane to the type II A string in $d=10$ \cite{Duff}. This opens up
an interesting possibility to interpret the spinning particles as low energy
effective models of the p-branes in higher-dimensional space-time, which
compact directions are associated with the spinning degrees of freedom of
the particle.

All this gives rise to the question about the construction of the mechanical
models of relativistic spinning particles in the space-time of arbitrary
dimension. The reach kinematical symmetries underlying the models enable one
to treat them as {\it elementary dynamical systems} in the Souriau sense
\cite{Souriau} and to apply for their description the full machinery of the
symplectic geometry. (For applications of this approach see also \cite{Duval}
). In the framework of this scheme the whole dynamical information about the
space-time and phase-space evolution of the system is encoded in a
presymplectic manifold ${\cal E}$ being a homogeneous transformation space
of the group $G$ for which the system is elementary one. For a given
physical system, the choice of ${\cal E}$ is very ambiguous and there is no
precise prescription for it. Fortunately, there is no matter how to choose
particular ${\cal E}$ when describing free particle: any ${\cal E}$ leads to
the proper classical dynamics \footnote{
For example, one can always identify ${\cal E}$ with the underlying symmetry
group $G$ itself.}.

Various models of spinning particles are known with $4D$ symmetry groups:
Poincar\'e \cite{Poincare,u} , de Sitter \cite{de Sitter} and Galilei \cite
{Galilei}. The $3D$ and $6D$ analogues to these models can be found in ref.
\cite{a,6D}, for superextensions see \cite{sms,sa}. The covariant
operatorial quantization of these models leads to the Hilbert space of
physical states carrying the unitary irreducible representation of the
respective groups. Some of these models have, however, a common problem of
constructing a consistent extension to the case of spinning particle subject
to exterior fields. The obstacle is that the space-time dynamics of the
particle, arising in these models in some cases, are characterized by
two-dimensional world-tubes rather then world-lines. The non-local behavior
of such a type, sometimes referred to as the phenomenon of {\it
Zitterbewegung} \cite{Z}, is usual for the relativistic particles with spin
and presents the main obstruction to the switching on a local interaction.
Note that the difficulty is not inherent to these systems, as they are, but
it is rather related to the way of their description, which basically
depends on the particular choice of ${\cal E}$ . In particular, the model of
the ref \cite{u} , in $d=4$, does not display the Zitterbewegung, and it
allows the consistent interaction.

In recent paper \cite{Integer} we have constructed the model for a massive
spinning particle living in $d$--dimensional space-time and coupled to an
arbitrary background of gravity and electromagnetism. The underlying
presymplectic manifold was identified with that for the spinless particle
times a regular (co)adjoint orbit of space rotations group
\begin{equation}
\begin{array}{c}
{\cal E}={\cal E}_{spinless}\times {\cal O}_{{\bf s}}\ , \\
\\
{\cal E}_{spinless}={\bf R}^{d-1,1}\times B\ ,\quad {\cal O}_{{\bf s}
}=SO(d-1)/[SO(2)]^r
\end{array}
\label{Os}
\end{equation}
Here $r=rank\ SO(d-1)=[(d-1)/2]$ and $B$ stands for the upper sheet of the
mass hyperboloid: $p_Ap^A=-m^2,\quad p_0>0$. The whole space of invariant
presymplectic structures on ${\cal E}$ is parametrized by $r+1$ numbers $m,\
{\bf s}=(s_1,s_2,...s_r)$ associated with mass and spin(s) of the particle.
The space-time motion of the particle is described here by the
one-dimensional world-lines (time-like geodesics of Minkowski space) and, as
a result, the model is free from the above mentioned obstacle to the
interaction. For the sake of explicit Poincar\'e covariance, the orbit $
{\cal O}_{{\bf s}}$ was symplectically embedded into $\oplus _{i=1}^r{\bf R}
_{{\bf C}}^{d-1,1}$ equipped with the natural action of the Lorentz group $
SO(d-1,1)$ and invariant symplectic form. This realization for ${\cal O}_{
{\bf s}}$ proves to be especially suitable for the covariant quantization of
integer-spin particle but it becomes inadequate when the half-integer spins
are considered since the quantum-mechanical description for the latter case
is based on the group $Spin(d-1,1)$ rather than $SO(d-1,1)$. In order to
take into account the half-integer spins, one should replace, from the very
beginning, the proper Lorentz group $SO(d-1,1)$ by its double covering $
Spin(d-1,1)$.

In this paper we modify the model in the way which provides a uniform
quantum-mechanical description for both integer and half-integer spins. For
these ends, the phase space of spinning degrees of freedom ${\cal O}_{{\bf s}
}$ is embedded into the carrier space of the Lorentz group ${\bf C}
^{2^{[d/2]}}\oplus _{i=1}^{r-1}{\bf R}_{{\bf C}}^{d-1,1}$ , where the first
factor transforms under the spinor representation. Then the (half-)integer
spin representations may be obtained by applying either geometric or Dirac
quantization to this model. A remarkable property of this realization for $
{\cal O}_{{\bf s}}$ is that the use of spinor variables makes possible to
resolve the mass-shell condition $p_Ap^A=-m^2$ in a Lorentz-invariant
manner. It can be thought about as an extension to the massive case of the
twistor realization known for the isotropic momenta of a massless particle
in special dimensions: 3, 4, 6 and 10. The distinction is that the spinor
variable carries now an information about both the mass hyperboloid $B$ and
the internal space for spin ${\cal O}_{{\bf s}}$. The details of this
construction are presented in the next section together with the
generalization to the case of minimal coupling to exterior gravitational and
electromagnetic fields.

In Section 3, we reformulate the model as Hamiltonian system with the first
and second class constraints and study the physical spectrum of the theory
within the Dirac quantization scheme. The physical wave functions, being
extracted by the quantum operators assigned to the constraints, are shown to
be in one-to-one correspondence with the Poincar\'e-irreducible
(spin-)tensor fields in Minkowski space.

We conclude the paper by discussing the obtained results and some further
perspectives.

\section{Classical description}

As it was mentioned in the previous section the extended phase space of the
massive spinning particle is constructed to be the direct product of the
extended phase space of the spinless particle and the manifold ${\cal O}_{
{\bf s}}$, responsible for spinning degrees of freedom. The former factor is
standardly embedded into the cotangent bundle $T^{*}({\bf R}^{d-1,1})$ of
the Minkowski space by the mass-shell condition
\begin{equation}
p_Ap^A+m^2=0  \label{ms}
\end{equation}
while the latter is identified with so-called flag manifod, which points may
be viewed as the sequences
\begin{eqnarray*}
0 &\subset &V_1\!\subset \!V_2\!\subset \!...\!\subset \!V_r\subset {\bf R}_{
{\bf C}}^{d-1,1}\ , \\
&& \\
~p &\perp &V_k\quad ,\ \qquad \dim V_k=k
\end{eqnarray*}
of complex $p$-transversal vector subspaces of complexified Minkowski space,
telescopically embedded into each other. In the previous paper \cite{Integer}
, we have suggested the holomorphic parametrization for ${\cal O}_{{\bf s}}$
with the help of $r$ independent complex vectors $Z_i^A\subset {\bf R}_{{\bf
C}}^{d-1,1},\ i=1,2,...r\,;\,\,A=0,1,...d-1$ subject to the conditions
\begin{equation}
(Z_i,Z_j)=0\ ,\quad (p,Z_i)=0\ ,\quad Z_i^A\sim Z_j^A\Lambda _i^j\quad ,
\label{par1}
\end{equation}
where $\Lambda _i^j$ is a complex non-degenerate upper-triangular $r\times r$
matrix, and $(...,...)$ denotes the inner product with respect to the
Minkowski metric $\eta _{AB}=diag(-,+\ldots ,+)$. In this realization, each $
V_k$ is spanned by the vectors $Z_1^A,Z_2^A,...Z_k^A$ and the last relation
in (\ref{par1}) establishes the equivalence between all such frames in $V_k$.

To construct a spinor realization for ${\cal O}_{{\bf s}}$, it is sufficient
to parametrize a subspace $V_r$ by a spinor variable. Since some subsequent
expressions may differ for the cases of even- and odd dimensions , the
formulae will be labeled with the letters $a$ and $b$ for the former and
latter case respectively.

Consider the Dirac spinor $\psi _a,\ a=1,...2^{[d/2]}$ subject to the
following constraints and equivalence relations
\begin{equation}
\begin{array}{c}
\psi _a\sim \lambda \psi _a\ ,\quad \lambda \in {\bf C}\setminus \{0\}\ , \\
\\
p_A\Gamma _a^{Ab}\psi _b=m\psi _a\ ,\quad Z_{kA}\Gamma _a^{Ab}\psi _b=0\
,\quad k=1,...r
\end{array}
\label{sp}
\end{equation}
Accounting (\ref{sp}) and the Fierz identities , the spinor bilinear is
decomposed in the basis of the Clifford algebra generated by $\Gamma $
-matrices as follows \addtocounter{equation}{1}:
$$
\psi \otimes \widetilde{\psi }=M_{A(r)}\{\Gamma ^{A(r)}-\frac{\displaystyle
(-1)^r}{\displaystyle m}p_B\Gamma ^{BA(r)}\}\ ,\eqno{(5.a)}
$$
$$
\psi \otimes \widetilde{\psi }=M_{A(r)}\Gamma ^{A(r)}\ ,\eqno{(5.b)}
$$
where $\widetilde{\psi }$ is the charge conjugated spinor\footnote{
The charge conjugation matrix $C$ is determined from the relation $\Gamma
_A^T=(-1)^rC\Gamma _AC^{-1}$} $\widetilde{\psi }{}^a=(\psi C)^a$. Hereafter
we use the shorthand notation $A(r)=A_1...A_r$. Tensor $M_{A(r)}=\frac{
(-1)^{r(r+1)/2}}{2^{[d/2]}r!}(\widetilde{\psi }\Gamma _{A(r)}\psi )$ obeys
equations
\begin{equation}
\begin{array}{c}
M_{[A(r)}Z_{kB]}=0\ ,\quad Z_k^AM_{AA(r-1)}=0\ ,\quad k=1,...r\ , \\
\\
p^AM_{AA(r-1)}=0\ ,\quad M_{A(r)}\sim \lambda ^2M_{A(r)}
\end{array}
\label{aa1}
\end{equation}
Making use of (\ref{aa1}) one can express $M_{A(r)}$ in terms of the $Z_i$ .
For these ends it is convenient to introduce another parametrization of $V_k$
with the help of $k$-forms ${\cal Z}^k$ determined from the equations
\begin{equation}
{\cal Z}^k(\bar Z_1,\bar Z_2,...\bar Z_k)=\det (Z_i\bar Z_j)\ ,\quad
i,j=1,...k\ ,
\end{equation}
that establish a relation between ${\cal Z}$ and $Z$
\begin{equation}
{\cal Z}_{A(k)}^k\sim Z_{1[A_1}Z_{2A_2}...Z_{kA_k]}  \label{con}
\end{equation}
Tensors ${\cal Z}_{A(k)}^k$ satisfy the following relations
\begin{equation}
\begin{array}{c}
{\cal Z}_{[A(i)}^i{\cal Z}_{B]B(k-1)}^k=0,\ \eta ^{AB}{\cal Z}_{AA(i-1)}^i
{\cal Z}_{BB(k-1)}^k=0,\ p^A{\cal Z}_{AA(k-1)}^k=0\ , \\
\\
{\cal Z}_{A(k)}^k\sim a_k{\cal Z}_{A(k)}^k\ ,\quad i,k=1,...r\ ,\quad i\geq
k\ ,\quad a_k\in {\bf C}\setminus \{0\}\ ,
\end{array}
\label{par2}
\end{equation}
resulting from the definitions (\ref{par1}). Now it is easy to see that the
general solution to the equations (\ref{aa1}) has the form
\begin{equation}
M_{A(r)}\sim {\cal Z}_{A(r)}^r
\end{equation}
Thus the $r$-dimensional complex vector subspace $V_r$ can be parametrized
by the spinor $\psi $ subject to the conditions (\ref{sp}). Making use of (
\ref{con}) we obtain the equivalent parametrization of ${\cal O}_{{\bf s}}$
in terms of $(r-1)$ complex vectors $Z_i^A,\ i=1,2,...r-1,\ A=0,1,...d-1$
and the spinor $\psi $ subject to equivalence relations
\begin{equation}
Z_i^A\sim Z_j^A\Lambda _i^j\ ,\quad \psi \sim \lambda \psi \ ,\quad \lambda
\in {\bf C}\setminus \{0\}  \label{eqrel}
\end{equation}
and constraints
\begin{equation}
(p,Z_i)=0\ ,\quad p_A\Gamma ^A\psi =m\psi \quad ,  \label{ptr}
\end{equation}
\begin{equation}
(Z_i,Z_j)=0\ ,\quad Z_{iA}\Gamma ^A\psi =0  \label{holonomic}
\end{equation}
Here $\Lambda _i^j$ is a complex non-degenerate upper-triangular $
(r-1)\times (r-1)$ matrix. In terms of introduced objects the most general
expression for the K\"ahler potential on ${\cal O}_{{\bf s}}$ is
\begin{equation}
\begin{array}{c}
\Phi =\frac{\displaystyle 1}{\displaystyle 2}\ln (\Delta _1^{s_1}...\Delta
_{r-1}^{s_{r-1}}\Delta _r^{s_r})\quad , \\
\\
\Delta _i=Z_{1A_1}...Z_{iA_i}\bar Z_1^{[A_1}...\bar Z_i^{A_i]}\ ,\quad
i=1,...,r-1\ ,\quad \Delta _r=(\bar \psi \psi )^2,
\end{array}
\end{equation}
where $\bar \psi ^a=({\psi ^{*}}\Gamma _0)^a$ is the Dirac conjugated spinor
. Note that $\Phi $ depends on $p$ implicitly in view of constraints (\ref
{ptr}). Under transformations (\ref{eqrel}) $\Phi $ changes to an additive
constant
\begin{equation}
\delta _{\Lambda ,\lambda }\Phi =\sum\limits_{k=1}^{r-1}\ln \left| \Lambda
_1^1\Lambda _2^2...\Lambda _k^k\right| ^{s_k}+\ln \left| \lambda \right|
^{2s_r}
\end{equation}

The direct product structure of ${\cal E}$ allows to introduce the one-form $
\theta $ being a sum of conventional one-form $p_Adx^A$ on ${\cal E}
_{spinless}$ describing the space-time dynamics of the particle and a
one-form on ${\cal O}_{{\bf s}}$ governing the spinning dynamics. We will
put
\begin{equation}  \label{aa3}
\theta =p_Adx^A+*d\Phi ,
\end{equation}
where the action of the star operator on the complex one-forms is defined as
$*(\alpha _Idz^I+\beta _Id\overline{z}^I)=-i(\alpha _Idz^I-\beta _Id
\overline{z}^I)$. Notice that $\theta $ is invariant under transformations (
\ref{eqrel}) modulo closed one-form and thus the Hamiltonian action for the
system may be chosen as
\begin{equation}  \label{act}
S=\int\limits_\gamma \theta
\end{equation}
The extremals of the action (\ref{act}) coincide with the leaves of $\ker
d\theta $. By construction $\ker d\theta $ is generated by the only vector
field ${\bf V}=p_A\partial /\partial x_A$ and hence the tangent vector to
the trajectory is proportional to ${\bf V}$. This means
\begin{equation}  \label{eqns}
\begin{array}{c}
\dot x^A=\mu p^A\ ,\quad \dot p_A=0\ , \\
\\
\dot Z_i^A=0\ ,\quad \dot \psi _a=0\ ,\quad (modulo\ transformations\ (\ref
{eqrel}))
\end{array}
\end{equation}
where $\mu $ is a Lagrange multiplier associated with the reparametrization
invariance. Thus the particle moves along the time-like geodesics in the
Minkowski space while the internal degrees of freedom do not evolve.

The interesting feature of the action (\ref{act}) is that the Dirac equation
on $\psi $ subject to the rest constraints (\ref{eqrel}, \ref{ptr}, \ref
{holonomic}) may be covariantly resolved with respect to $p_A$. To solve
this equation, we multiply it by $\psi _c$ and make use of the decomposition
(5) together with Fierz identities for spinor bilinears. This results in
relations \addtocounter{equation}{2}
$$
p^A(\widetilde{\psi }\Gamma _{AA(r-1)}\psi )=0,\ (\widetilde{\psi }\Gamma
_{AA(r)}\psi )=\frac{\displaystyle (-1)^{r+1}(r+1)}{\displaystyle m}p_{[A}(
\widetilde{\psi }\Gamma _{A(r)}\psi )\ ,\eqno{(19.a)}
$$
$$
p^A(\widetilde{\psi }\Gamma _{AA(r-1)}\psi )=0,\ \epsilon _{A(r)BC(r)}p^B(
\widetilde{\psi }\Gamma ^{C(r)}\psi )=(-1)^{1+r(r+1)/2}r!mi^r(\widetilde{
\psi }\Gamma _{A(r)}\psi )\eqno{(19.b)}
$$
Contracting the second equation in (19.a) with $(\widetilde{\psi }\Gamma
^{A(r)}\psi )^{*}$ one can express $p_A$ via the spinor variables
$$
p_A=\bar p_A(\psi ,\psi ^{*})\equiv (-1)^{r+1}m\frac{[(\widetilde{\psi }
\Gamma _{AB(r)}\psi )(\widetilde{\psi }\Gamma ^{B(r)}\psi )^{*}+c.c.]}{
2\left| (\widetilde{\psi }\Gamma _{C(r)}\psi )\right| ^2}\eqno{(20.a)}
$$
For the odd-dimensional case the similar trick (with $(\widetilde{\psi }
\Gamma ^{B(r)}\psi )^{*}$ replaced by $\epsilon ^{A(r)CD(r)}(\widetilde{\psi
}\Gamma _{D(r)}\psi )^{*}$ ) yields
$$
p_A=\bar p_A(\psi ,\psi ^{*})\equiv m\frac{(-1)^{r(r-1)/2}i^r\epsilon
_{AB(r)C(r)}(\widetilde{\psi }\Gamma ^{B(r)}\psi )(\widetilde{\psi }\Gamma
^{C(r)}\psi )^{*}}{r!\left| (\widetilde{\psi }\Gamma _{D(r)}\psi )\right| ^2}
\eqno{(20.b)}
$$
Since equations (19) form the overdetermined system, the expressions (20)
being inserted back into (19) will lead to some consistency conditions which
should be imposed on $\psi $ besides the holonomic constraints (\ref
{holonomic}). Note that the mass-shell condition (\ref{ms}) reduces to the
purely algebraic relation on $\psi $ which is valid due to these
constraints. Thus we obtain the parametrization of the massive hyperboloid
in terms of the spinor $\psi _a$ subject to the conditions (\ref{eqrel}, \ref
{holonomic}) and (19) (with $p$ replaced by $\bar p$)\footnote{
These constraints will constitute spin-shell conditions, i.e. they will fix
the value of the phase space Casimir functions of the Poincare' group.}.
This resembles to the twistor parametrization of the light-cone \cite
{Cederwall} which, however, is connected with the division algebras and
therefore exists in $3$, $4$, $6$ and $10$ dimensions only.  The derived
construction may be considered as its generalization to the case of
arbitrary dimensional space-time with the exception that the spinor $\psi $
contains information about both the space-time momentum and intrinsic
degrees of freedom. For  the special case of $d=4$ such a consruction was
previosly used in \cite{Biedenharn}.

Substituting $\bar p_A$ in (\ref{aa3}), we immediately derive the Lagrangian
of the system
\begin{equation}
{\cal L}=\bar p_A(\psi ,\psi ^{*})\dot x^A-i(\dot Z_i^A\partial _A^i\Phi
+\dot \psi _a\partial ^a\Phi -c.c.)  \label{lagr}
\end{equation}
The Lagrangian is obviously invariant under the global Poincar\'e
transformations, reparametrizations of the worldline and changes to a total
derivative under the gauge transformations associated with equivalence
relations (\ref{eqrel})
\begin{equation}
Z_i^{\prime A}(\tau )=Z_j^A(\tau )\Lambda _i^j(\tau )\ ,\quad \psi
_a^{\prime }(\tau )=\lambda (\tau )\psi _a(\tau )  \label{gauge}
\end{equation}
The global Poincar\'e symmetry leads to the on-shell conservation of the
Hamiltonian counterparts of Poincar\'e generators ${\bf P}_A,\ {\bf M}_{AB}$
\begin{equation}
\begin{array}{c}
{\bf P}_A=\bar p_A\ ,\quad {\bf M}^{AB}=x^A\bar p^B-x^B\bar p^A+S^{AB}\ , \\
\\
S^{AB}=2i\{Z_i^A\partial ^{iB}-\overline{Z}_i^A\overline{\partial }
^{iB}-(\Sigma ^{AB})_b^{\ a}(\psi _a\partial ^b+\overline{\psi }_a\overline{
\partial }^b)\}\Phi
\end{array}
\end{equation}
$(\Sigma _{AB})_a^{\ b}=-\frac 14[\Gamma _A,\Gamma _B]_a^{\ b}$ being the
Lorentz generators in the spinor representation.

Let us now secify the expressions (19, 20) to the case of $d=4$. Then the
spinning sector is parametrized by one variable $\psi $ defined modulo
multiplication by a complex nonzero constant and subject to the Dirac
equation. The expression for $p_A$ reads
\begin{equation}
p_A=\bar p_A(\psi ,\psi ^{*})\equiv m\frac{(\widetilde{\psi }\Gamma
_{AB}\psi )(\widetilde{\psi }\Gamma ^B\psi )^{*}+c.c.}{2\left| (\widetilde{
\psi }\Gamma _C\psi )\right| ^2}  \label{d4p}
\end{equation}
and the consistency conditions take the form
\begin{equation}
\bar p^A(\widetilde{\psi }\Gamma _A\psi )=0\ ,\quad (\widetilde{\psi }\Gamma
_{AB}\psi )=\frac{\displaystyle 2}{\displaystyle m}\bar p_{[A}(\widetilde{
\psi }\Gamma _{B]}\psi )  \label{d4c}
\end{equation}
The independent equations may be easily picked out among (\ref{d4c}) after
transition to the two-component spinor formalism, $\psi =\binom \lambda
{\bar \chi }$.  In terms of the Weyl spinors the Lagrangian can be rewritten
as
\begin{equation}
{\cal L}=m\frac{{\dot x}_A(\sigma _{a\dot a}^A)(\lambda ^a\bar \chi ^{\dot
a}+\chi ^a\bar \lambda ^{\dot a})}{2(\lambda _a\chi ^a)}+is\frac{({\dot
\lambda }_a\chi ^a)-({\dot{\bar \lambda} }_a\bar \chi ^a)}{({\lambda }_a{\chi }^a)}
\end{equation}
and the spinors $\lambda $ and $\chi $ are restricted to satisfy
\begin{equation}
{\bf Im}(\lambda _a\chi ^a)=0  \label{sphol}
\end{equation}
As is seen, the momentum conjugated to $x^A$ is the function of $\lambda $
and $\chi $ and the mass-shell condition is valid due to the (\ref{sphol}).

The action (\ref{act}) allows the coupling to the background gravitational
and electromagnetic fields. The standard method of introducing the
interaction implies to consider the gauge fields associated with the
vielbein $e_\mu ^A$ , torsion-free spin connection $\omega _{\mu AB}$ and
the electromagnetic potential $A_\mu $. Then the minimal covariantization of
(\ref{act}) reads
\begin{equation}
\begin{array}{c}
S=\int \widetilde{\theta }\ , \\
\\
\widetilde{\theta }=(p_Ae_\mu ^A-eA_\mu )dx^\mu +*D\Phi ,
\end{array}
\label{crvdact}
\end{equation}
where $D$ is the Lorentz covariant differential along the particle worldline
\begin{equation}
\begin{array}{c}
DZ_i^A=dZ_i^A+dx^\mu \omega _\mu {}^A{}_BZ_i^B\ , \\
\\
D\psi _a=d\psi _a+dx^\mu \omega _{\mu AB}\Sigma _a^{ABb}\psi _b,
\end{array}
\end{equation}
and $e$ is the electric charge. The relations (\ref{ms}, \ref{ptr}) on the
momentum $p_A$ are still assumed to hold. The action (\ref{crvdact})
generates the following equations of motion
\begin{equation}
\begin{array}{c}
\dot x^\mu =\lambda e_A^\mu p^A\ ,\quad \frac{\displaystyle Dp_A}{
\displaystyle d\tau }+eF_{AB}p^B=(\lambda /4)R_{ABCD}p^BS^{CD}\ , \\
\\
\frac{\displaystyle DZ_i^A}{\displaystyle d\tau }=0\ ,\quad \frac{
\displaystyle D\psi _a}{\displaystyle d\tau }=0
\end{array}
\label{crvdeqns}
\end{equation}
Here $F_{\mu \nu }$ is the strength tensor of the electromagnetic field and $
R_{\alpha \beta \gamma \delta }$ is a curvature of the space-time. Thus the
dynamics of the particle in the curved space-time is described by first two
equations while the motion in the spinning sector reduces to the parallel
transport along the worldline.

To conclude, let us note that the gauge symmetry (\ref{gauge}), being
required to survive on the quantum level, implies the restriction on the
possible values of the parameters $s_i$ entering the K\"ahler potential.
Proceeding by analogy with the integer spin case \cite{Integer} one can show
that in quantum theory $s_i,\ i=1,...r-1$ are constrained to be integer and $
s_r$ - (half-)integer numbers.

\section{Quantization}

In this section we will present the covariant quantization of the model
described. Let us start with the Lagrangian (\ref{lagr}) complemented with
the conditions (\ref{holonomic}) and (19). This Lagrangian leads to the
following primary constraints
\begin{equation}
\begin{array}{c}
T_A=p_A-\bar p_A\approx 0\ , \\
\\
\nabla _A^i=q_A^i+i\partial _A^i\Phi \approx 0\ ,\quad \bar \nabla
_A^i=(\nabla _A^i)^{*}\ , \\
\\
\nabla ^a=\pi ^a+i\partial ^a\Phi \approx 0\ ,\quad \bar \nabla ^{\dot
a}=(\nabla ^a)^{*}
\end{array}
\label{q1}
\end{equation}
(as well as the above mentioned holonomic ones). Here $p_A$, $q_A^i$ and $
\pi ^a$ are the momenta conjugated to $x^A$, $Z_i^A$ and $\psi _a$
respectively, $\bar p_A$ is defined by rels. (20). The system (\ref{q1})
allows a transition to the more suitable basis of constraints where
relations (19) have been already accounted
\begin{equation}
\begin{array}{c}
T_i=(p,Z_i)\approx 0\ ,\quad T_a=p_A\Gamma ^A\psi -m\psi \approx 0\ , \\
\\
\bar \nabla _A^i\approx 0\ ,\quad \bar \nabla ^{\dot a}\approx 0
\end{array}
\label{q2}
\end{equation}
and the complex conjugated constraints are also implied. Stabilization of
the constraints yields the mass-shell condition
\begin{equation}
T=p_Ap^A+m^2\approx 0  \label{q3}
\end{equation}
It should be noted that the constraints (\ref{q2}, \ref{q3}) are not
independent. Moreover, only first class constraints amongst (\ref{q2})
\footnote{
Of course, (\ref{q3}) is also first class.} can be covariantly extracted:
\begin{equation}
\begin{array}{c}
\Pi _j^i=(Z_j,q^i)+i\delta _j^il_i\ ,\quad \bar \Pi _j^i=(\Pi _j^i)^{*}\
,\quad i\geq j\ , \\
\\
\Pi =(\psi _a\pi ^a)+il_r\ ,\quad \bar \Pi =(\Pi )^{*}\ , \\
\\
l_i=\sum\limits_{k=i}^{r-1}s_k\ ,\quad l_r=2s_r
\end{array}
\end{equation}
Constraits $\Pi _j^i$ and $\Pi $ generate the gauge transformations (\ref
{eqrel}). Following the covariant quantization scheme, the set of variables $
(x,p,Z,q,\psi ,\pi )$ is associated to the self-adjoint operators acting in
a Hilbert space of the particle states. The physical states $|\Psi \rangle $
are singled out from the space of smooth functions on ${\bf R}^{d-1,1}\times
{\bf C}^{(r-1)d}\times {\bf C}^{2^{d/2}}$ by imposing the first class
constraint operators and a half of the second class ones. Since the latter
cannot be presented explicitly we will impose on the physical states the
operatorial counterparts of the expressions (\ref{q2}) (accounting thereby
the half of the second class constraints), together with $\widehat{\bar \Pi
}{}_j^i$ and $\widehat{\bar \Pi }$. As a result, the physical states are
annihilated by the operators $\widehat{T}$, $\widehat{T}_a$, $\widehat{T}_i$
, $\widehat{\Pi }$, $\widehat{\Pi }{}_j^i$, $\widehat{\bar \nabla }{}^{\dot
a}$, $\widehat{\bar \nabla }{}_A^i$. In the coordinate representation for $
(Z,q,\psi ,\pi )$ and the momentum one for $(x,p)$
\begin{equation}
q_A^i\rightarrow -i\partial _A^i\ ,\quad \pi ^a\rightarrow -i\partial ^a\
,\quad x^A\rightarrow i\partial ^A
\end{equation}
the constraint equations for the wave function $\Psi (Z,\overline{Z},\psi
,\psi ^{*},p)$ take the following explicit form:
\begin{equation}
(Z_j^A\partial _A^i-\delta _j^in_i)\Psi =0\ ,\quad (\psi _a\partial
^a-n)\Psi =0  \label{q4}
\end{equation}
\begin{equation}
\bar \partial _A^i\Psi +(\bar \partial _A^i\Phi )\Psi =0\ ,\quad \bar
\partial ^{\dot a}\Psi +(\bar \partial ^{\dot a}\Phi )\Psi =0  \label{q5}
\end{equation}
The wave function is defined on the surface (\ref{ptr}, \ref{holonomic}).
Note that the constants $n$, $n_i$ may differ from their classical values $l$
, $l_i$ due to the different operator ordering prescriptions for $\widehat{
\Pi }$ and $\widehat{\Pi }{}_j^i$. We fix the ambiguity in the factor
ordering by the requirement that the physical wave functions should remain
unchanged under the gauge transformations (\ref{eqrel}) which implies
vanishing of $n$ and $n_i$. Any other ordering will lead to the unitary
equivalent theory. The general solution for (\ref{q5}) reads
\begin{equation}
\Psi (Z,\overline{Z},\psi ,\psi ^{*},p)=\exp (-\Phi (Z,\overline{Z},\psi
,\psi ^{*}))\Theta (Z,\psi ,p)  \label{q6}
\end{equation}
Substituting (\ref{q6}) in (\ref{q4}) one gets
\begin{equation}
(Z_j^A\partial _A^i-\delta _j^il_i)\Theta =0\ ,\quad (\psi _a\partial
^a-l_r)\Theta =0  \label{q7}
\end{equation}
Since the manifold ${\cal O}_{{\bf s}}$ is compact the eigenvalues $l_i$, $
l_r$ prove to be integers with eigenfunctions $\Theta $ represented by the
polynomials of the form
\begin{equation}
\Theta (Z,\psi ,p)=\Theta
(p)_{A(l_1)B(l_2)...C(l_{r-1})a(l_r)}Z_1^{A(l_1)}Z_2^{B(l_2)}...Z_{r-1}^{C(l_{r-1})}
\widetilde{\psi }^{a(l_r)}\ ,
\end{equation}
where we denote $Z_1^{A(l)}\equiv Z_1^{A_1}\cdots Z_1^{A_l}$, $\widetilde{
\psi }{}^{a(l)}\equiv \widetilde{\psi }{}^{a_1}...\widetilde{\psi }{}^{a_l}$
; the symmetry of the indices is described by the Young diagram

\unitlength=0.63mm \special{em:linewidth 0.4pt} \linethickness{0.4pt}
\begin{picture}(30.00,40.00)(-50.0,117.0)
\put(10.00,150.00){\line(1,0){64.00}}
\put(74.00,150.00){\line(0,-1){8.00}}
\put(74.00,142.00){\line(-1,0){64.00}}
\put(10.00,142.00){\line(0,1){8.00}}
\put(10.00,142.00){\line(0,-1){8.00}}
\put(10.00,134.00){\line(1,0){50.00}}
\put(60.00,134.00){\line(0,1){8.00}}
\put(10.00,128.00){\line(1,0){42.50}}
\put(52.50,128.00){\line(0,-1){8.00}}
\put(52.50,120.00){\line(-1,0){42.50}}
\put(10.00,120.00){\line(0,1){8.00}}
\put(18.00,150.00){\line(0,-1){16.00}}
\put(66.00,150.00){\line(0,-1){8.00}}
\put(52.00,142.00){\line(0,-1){8.00}}
\put(38.50,128.00){\line(0,-1){8.00}}
\put(18.00,128.00){\line(0,-1){8.00}}
\put(26.00,150.00){\line(0,-1){16.00}}
\put(26.00,128.00){\line(0,-1){8.00}}
\put(14.00,146.00){\makebox(0,0)[cc]{$A_1$}}
\put(14.00,138.00){\makebox(0,0)[cc]{$B_1$}}
\put(14.00,124.00){\makebox(0,0)[cc]{$C_1$}}
\put(22.00,146.00){\makebox(0,0)[cc]{$A_2$}}
\put(22.00,138.00){\makebox(0,0)[cc]{$B_2$}}
\put(22.00,124.00){\makebox(0,0)[cc]{$C_2$}}
\put(46.00,124.00){\makebox(0,0)[cc]{$C_{l_{r-1}}$}}
\put(56.00,138.00){\makebox(0,0)[cc]{$B_{l_2}$}}
\put(70.00,146.00){\makebox(0,0)[cc]{$A_{l_1}$}}
\put(32.50,146.00){\makebox(0,0)[cc]{$\cdots$}}
\put(32.50,138.00){\makebox(0,0)[cc]{$\cdots$}}
\put(32.50,124.00){\makebox(0,0)[cc]{$\cdots$}}
\put(32.50,131.00){\makebox(0,0)[cc]{$\cdots$}}
\end{picture}

Notice that, by virtue of relations (\ref{eqrel}, \ref{ptr}, \ref{holonomic}
), the coefficients $\Theta (p)_{A...Ca...}$ are assumed to be $p$
-transversal, traceless and $\Gamma $-traceless
\begin{equation}
p^A\Theta (p)_{...A...}=0\ ,\quad \eta ^{AB}\Theta (p)_{...A...B...}=0\
,\quad \Gamma _a^{Ab}\Theta (p)_{...A...b...}=0  \label{q8}
\end{equation}
In spinor indices $\Theta $ is subject to the Dirac equation
\begin{equation}
p_A\Gamma _a^{Ab}\Theta (p)_{...b...}=m\Theta (p)_{...a...}  \label{q9}
\end{equation}
Equations (\ref{q8}, \ref{q9}) constitute together the full set of $d$
-dimensional relativistic wave equations on irreducible massive spin tensor
fields which are obtained after Fouriau transform of $\Theta (p)$. The space
${\cal H}_{m,{\bf s}}$ of functions $\Psi $ (\ref{q6}) representing the
particle states is endowed with the Hilbert space structure with respect to
the following Hermitian inner product
\begin{equation}
<\Psi _1|\Psi _2>=\int\limits_{p^2=-m^2}\frac{d{\bf p}}{p_0}\int\limits_{
{\cal O}_{{\bf s}}}d\mu \bar \Psi _1\Psi _2
\end{equation}
where
\[
d\mu =d*d\Phi \wedge d*d\Phi \wedge ...\wedge d*d\Phi
\]
is the Liouville measure on ${\cal O}_{{\bf s}}$. Integration over the
spinning degrees of freedom may be performed explicitly resulting with the
standard field-theoretical inner product
\begin{equation}
<\Psi _1|\Psi _2>=N\int\limits_{p^2=-m^2}\frac{d{\bf p}}{p_0}\overline{
\Theta }(p)_1^{A(l_1)B(l_2)...C(l_{r-1})a(l_r)}\Theta
(p)_{2A(l_1)B(l_2)...C(l_{r-1})a(l_r)}
\end{equation}
where $\overline{\Theta }^{...a(l_r)}=\Theta ^{*...}{}_{\dot a(l_r)}(\Gamma
_0)^{\dot a_1a_1}...(\Gamma _0)^{\dot a_{l_r}a_{l_r}}$ and $N$ is a constant
depending on spin of the particle.

Thus the covariant operatorial quantization of the model yields the
irreducible representation of the Poincar\'e group with quantum numbers
fixed by the constants originally entering the K\"ahler potential. We would
like also to note that the procedure of geometric quantization provides
another way of quantizing these systems with the same result (see \cite
{Integer} where the model of integer spin massive particle is quantized
within this approach).

\section{Conclusion}

Let us discuss some possible links between the spinning particle model
suggested and some problems of interest in other topics of the field theory
and mention some open questions related to the paper results.

In this paper we have extended the description of the integer spin particle
\cite{Integer} to a general (half-)integer case in arbitrary dimension. The
key ingredients of the construction are the covariant realization of the
spin phase space ${\cal O}_{{\bf s}}$ by means spinor variables subject to
certain covariant constraints and the respective invariant one-form (\ref
{aa3}), which defines the action functional for the model.

It is curious to note that, due to the constraints, the particle momentum
can be explicitly expressed in terms of the spinning variables that
resolves, in fact, the mass-shell condition. The corresponding Lagrangian
(\ref{lagr}), being written without canonical momenta $p_A$, leads to the
Hamiltonian formulation, where the spin-shell conditions include an actual
first class constraint, related to the reparametrization invariance, whereas
the mass-shell condition is an identity on the second class constraint
surface. In this sense, we observe an interchange of the common roles of the
spin- and mass-shell conditions in spinning particle models. This unusual
situation could be useful for constructing action functionals for the
corresponding higher spin fields. For these fields, being defined on the
mass-shell, the spin-shell conditions would serve as field equations. One
may hope that the longstanding problem of the higher spin interaction would
have more transparent form in this formulation, as just the spinning
structure of the fields is in the focus of these equations, whereas
inessential mass-shell is obeyed by construction.

As far as the interaction problem is concerned, we may mention that, at
least at the formal level, it is solved in this paper for the coupling of
the spinning particle to a general gravity and electromagnetic background
fields. The extension can be immediately got to the particle interaction
with the dynamical fields by adding the free field Lagrangian in the action.
However, the less formal aspect related to the selfacceleration problem \cite
{vH} remains unclear now in the selfconsitent theory of the particle coupled
to the dynamical fields. In ref. \cite{vH} has been shown that the
selfacceleration of the particle does not occur when it is coupled to a very
special spectrum of the fields. This special field spectrum is known for a $
d=4$ spinless particle only. For higher dimensions and/or nonzero spin, the
solution to the selfacceleration problem could be different. It should be
mentioned also that the interaction to the non-abelian gauge fields requires
to equip the particle with isospinning degrees of freedom. This could be
done, in general, along the same lines as it is performed for a genuine spin
in this paper, although the isospin requires different inner manifold
instead of ${\cal O}_{{\bf s}}$ (\ref{Os}). In this way, the isopin-shell
conditions should appear as phase space constraints. However, it is unclear
from the outset, whether the interplay between spin and isospin in the
constraint algebra would be consistent to an arbitrary Yang-Mills background
field.

Another problem, where this model may seem to gain some importance, is the
relationship between strings and spinning particles. This relationship is
commonly known at the level of the quantum string state spectrum which
includes an infinite number of massive excitations of various spins. These
excitations are usually thought about as states of certain spinning
particles. However, it remains yet unclear how the nonzero string modes
(subject to the Virasoro constraints) may form the spinning sector of the
particle phase space. In other words, the question is how the surface of
Virasoro constraints in the string phase space is stratified into the
spinning particle presymplectic manifolds ${\cal E}$ (\ref{Os}).
Comprehension of the structure of this stratification seems to be relevant
for study of a reduction procedure in string theory. The geometry of the
spinning particle phase space, revealed in this paper, forms a basis to
study the relationship between strings and particles in this direction.

\hspace{-0.5cm}

\section{Acknowledgments}

This work is partially supported by the grant Joint INTAS-RFBR 95-829 and
RFBR 98-02-16261. S.L.Lyakhovich appreciates the financial support from the
ISSEP.

\end{document}